\documentstyle[aps,prb,preprint]{revtex}
\title{Anomalous Raman shift in the ternary fullerides with $t_{1g}$ states}
\author{X. H. Chen}
\address{Japan Advanced Institute of Science  and Technology \\
Tatsunokuchi, Ishikawa  923-1292, Japan\\
and Department of Physics, University of Science and Technology of China\\
Hefei, Anhui 230026, P. R. China}
\author{T. Takenobu, T. Muro,  H. Fudo,  and Y. Iwasa}
\address{Japan  Advanced Institute  of  Science  and Technology\\
Tatsunokuchi, Ishikawa 923-1292,  Japan}
\date{February 3, 1999}
\begin{document}
\maketitle
\begin{abstract}

Raman spectra have been studied on two kinds of highly doped  fullerides
 of $A_xBa_3C_{60}$ (x=0, 3; A=K, Rb) and $K_xSm_{2.75}C_{60}$ 
(x=0, 3.25). It was found that  the Raman  spectra  are  essentially  identical
to each other for all ternary fullerides. The results show a crossover point
at the boundary of $t_{1u}$ and $t_{1g}$ bands for the charge transfer
 dependence of  Raman shift. Particularly, the totally symmetric  $A_g(2)$
mode  in $t_{1g}$  fullerides  cannot  be understood by  a simple 
extrapolation  from the low doped $t_{1u}$  fullerides, where the $A_g(2)$ 
mode follows a characteristic shift  of $\sim6.3 cm^{-1}$ per elementary 
charge . The present  result  shows that the Raman spectra of $t_{1g}$
states cannot be explained  only by charge transfer.

\end{abstract}
 
\vspace{5mm}


\vspace{5mm}

Raman scattering is a useful technique to study the vibrational properties of
the $C_{60}$ molecule and its doped derivative compounds. Raman  scattering 
is widely used to evaluate the electron-phonon  coupling constant $\lambda$
for the  doped $C_{60}$ superconductors  basing on the  analysis  of the 
linewidths.\cite{mitch,zhou,winter} In addition,   the tangential  pinch 
$A_g(2)$ mode is of particular interest in the studies of the doped  $C_{60}$  
as it yields a strong and  narrow line in the Raman spectrum. The different 
stages of doping can be easily  followed by in-situ Raman-scattering
experiments. A continuous  change  of  line intensity for the $A_g(2)$  mode of
the stable phase is observed and the  doping process leads to a characteristic  
downshift of this line regarding the  number of electrons transferred to the
 $C_{60}$  molecule. A down-shift of 6-7 $cm^{-1}$  per elementary charge 
on $C_{60}$ independent   of the doping ion is observed.
\cite{haddon,duclos,pichler,wang,kuzmany}  The  $AC_{60}$  
phase was discovered  by down-shift  of the $A_g$ pinch mode in in-situ  
Raman-scattering experiments.\cite{winter1}

 $A_3Ba_3C_{60}$  phases (A=K, Rb, Cs) with a formal 
$C_{60}^{-9}$ charge  and half-filling of the $t_{1g}$  are recently  
reported.\cite{iwasa}  As the intercalation  host,  $Ba_3C_{60}$   is a 
vacancy-ordered derivative of the bcc $A_6C_{60}$  structure  with half of the 
cation sites empty  ($A15$ structure). Three alkali metal cations are 
introduced into $Ba_3C_{60}$  to form a cation-disordered $A_3Ba_3C_{60}$  
phase  isostructural  with $A_6C_{60}$.  $K_3Ba_3C_{60}$  with half-filled 
$t_{1g}$  band exhibits superconductivity  at  5.6 K. However, the insertion of 
large $A^+$  cations  leads  to a decrease in $T_c$, contrary to the behavior  
of the  $A_3C_{60}$ phases.\cite{iwasa1}

To investigate the vibrational properties of ternary  fullerides  and comparison
of physical properties  in between $t_{1u}$ and $t_{1g}$  fullerides, we have
carried  out a Raman  scattering study on the fullerides $A_xBa_3C_{60}$ (x=0, 3;
A=K, Rb) and $K_xSm_{2.75}C_{60}$  (x=0, 3.25). The  results  show that the
Raman spectra are amazingly similar  to each other for all ternary  fullerides
$A_3Ba_3C_{60}$ and $K_{3.25}Sm_{2.75}C_{60}$. An anomalous  Raman shift of the
two $A_g$ modes is observed when alkali metals are introduced into $Ba_3C_{60}$
and $Sm_{2.75}C_{60}$. It does not follow the characteristic relation between
charge transfer  and Raman  shift widely observed for the two $A_g$ modes  in
alkali  and Ca-doped  $C_{60}$.\cite{haddon,duclos,pichler,wang,chen}

Samples of $Ba_3C_{60}$   and $Sm_{2.75}C_{60}$  were synthesized by  
reacting stoichiometric  amount of powers of Ba, Sm, and $C_{60}$.  A quartz  
tube with mixed powder inside was sealed under high vacuum of about  
$2\times 10^{-6}$ torr, and heated  at $550\sim600$  $^oC $ for three  days.  
Synthesis of  $A_3Ba_3C_{60}$  and $K_{3.25}Sm_{2.75}C_{60}$ was carried out 
in a similar manner to that of alkali doping into pure $C_{60}$,  A piece  
of alkali and 
$Ba_3C_{60}$  or $Sm_{2.75}C_{60}$  powders were loaded in a Pyrex tube, 
which was sealed under   $2\times 10^{-6}$ torr  and calcined at 250 $^oC$  for 
three days.  X-ray diffraction analysis was performed by a system equipped with 
a 4.5 kW rotating molybdenum anode  as the x-ray generator  and an imaging plate
 (IP, MAC, Science , DIP320V ) as the detector.
X-ray  diffraction showed that  all samples  were single phase, which was also 
confirmed by the single peak feature of the pentagonal  pinch  $A_g(2)$  mode  
in the Raman spectra.

Raman scattering  experiments  were carried out  using the  632.8 nm line of
a He-Ne laser in the Brewster angle backscattering geometry. The scattering
light  was detected with a Dilor xy multichannel spectrometer using a spectral
resolution of  3 $cm^{-1}$. In order to obtain  good  Raman
spectra, the samples  were ground  and pressed  into  pellets  with pressure
of about  20 $kg/cm^2$,  which  were sealed  in Pyrex tubes  under a
high vacuum of $10^{-6}$  torr.

As reported in reference  10, X-ray diffraction shows that the structure is
changed from $A15$ phase to a cation-disordered $A_3Ba_3C_{60}$ phase  
 isostructural with bcc  $A_6C_{60}$  (a=K, Ba) when alkali metal  is
 intercalated into  $Ba_3C_{60}$. Figure 1 shows the X-ray powder 
diffraction patterns  for the samples  $Sm_{2.75}C_{60}$ and 
$K_{3.25}Sm_{2.75}C_{60}$. For the pattern of $Sm_{2.75}C_{60}$, all peaks
can be  indexed with an orthorhombic lattice parameters  a=28.158 \r{A},
b=28.077 \r{A}, and c=28.270 \r{A}, which is consistent with previous
report.\cite{chen1}   It is easily seen in Fig.1 that the diffraction peak is
much less in  the pattern of $K_{3.25}Sm_{2.75}C_{60}$ than in the  pattern
of  $Sm_{2.75}C_{60}$. In addition, no peak is observed at $2 \theta$  below
5 degree. It suggests that  the structure  of $K_{3.25}Sm_{2.75}C_{60}$ 
should be simple  relative to that of $Sm_{2.75}C_{60}$. We  can index  all  
 diffraction  peaks  using a body-center ed-cubic   cell  with lattice parameter
  a=11.093\ \r{A} except  for the two peaks  marked by arrows in Fig.1. 
It is known that  the  basic structure  of $Sm_{2.75}C_{60}$  is fcc, but  the 
 cation vacancy  ordering  in tetrahedral sites  leads  to  a superstructure  
accompanied with a slight lattice deformation from cubic to orthorhombic.
The Sm cations  occupying the octahedral and tetrahedral sites experience 
off-center  displacements  since  one out of every eight tetrahedral sites in the 
subcell is vacant.\cite{chen1}  X-ray diffraction suggests that  introduction of
the alkali cations  into $Sm_{2.75}C_{60}$  leads  to the disappearance of  
cation-vacancy   ordering  and  formation  of  a cation-disordered   phase with 
composition $K_{3.25}Sm_{2.75}C_{60}$, similarly
 to the case of $A_3Ba_3C_{60}$.   A detailed  analysis  on the structure  of  
 $K_{3.25}Sm_{2.75}C_{60}$  will be reported  elsewhere. 

Figure 2 shows Raman spectra of $Ba_3C_{60}$,  $K_3Ba_3C_{60}$,  and 
$Rb_3Ba_3C_{60}$. The positions ( $\omega$  ) and halfwidths ( $\gamma$ ) 
of the Raman modes observed are  listed in Table  I for all samples.  In the  
$Ba_3C_{60}$  spectrum, there  are about  13 strong Raman lines observed, 
some of which are doublets.  The low-and  high-frequency  $A_g$  derived modes  
are located at  506 and 1430.8 $cm^{-1}$, respectively. The position of $ A_g(2)$  
pentagonal pinch mode  is identical to that of $K_6C_{60}$, suggesting  that  the 
charge transfer from Ba to $C_{60}$  is complete,  also being  consistent  with
$\sim6.3 cm^{-1}$  redshift per electron
 relative to neutral $C_{60}$.\cite{kuzmany} However,  the up-shift of 13 
$cm^{-1}$ for the  radial  $A_g(1)$ mode  is larger than  that  for $K_6C_{60}$
(9 $cm^{-1}$ ).\cite{zhou1}  It is to be pointed out that the Raman spectrum of
 $Ba_3C_{60}$ is amazingly similar to that of $K_6C_{60}$ except   for
the relative intensities  between $A_g(2)$  and $H_g(2)$ modes.

In the case of $K_3Ba_3C_{60}$   and $Rb_3Ba_3C_{60}$,   the similar Raman
 lines to $Ba_3C_{60}$   are observed. However, the strongest line is $A_g(2)$ 
mode, similarly to the case of $K_6C_{60}$  with bcc structure. It is worth 
pointing out that  the $K_3Ba_3C_{60}$  and $Rb_3Ba_3C_{60}$  spectra 
are essentially identical to each other. These spectra are relatively insensitive
 to the  choice  of alkali-metal ion species.  This behavior  is similar to the
case of $A_3C_{60}$ and $A_6C_{60}$ ( A=K, Rb ) compounds,  where this 
behavior  is explained  both  by a weak coupling between  $C_{60}$  and 
alkali  cations  and by  a complete charge transfer  from the alkali-metals
 to $C_{60}$ molecules.\cite{zhou1}   However, an  anomalous  Raman 
shift for the two $A_g$-derived  modes  is observed in $K_3Ba_3C_{60}$  and  
$Rb_3Ba_3C_{60}$.  The first thing to be noted is the  $A_g(2)$ mode  observed
 at  1425.5 ( 1424.1 ) $cm^{-1}$   in $K_3Ba_3C_{60}$   ($Rb_3Ba_3C_{60}$).  
The relative down-shift of $A_g(2)$ mode measured from that of $Ba_3C_{60}$ 
  is only 5.3 and 6.7  $cm^{-1}$  for $K_3Ba_3C_{60}$   and  
$Rb_3Ba_3C_{60}$ ,  respectively.  These values are much smaller  than that 
expected basing on the established empirical relation between the formal $C_{60}$ 
valence and the Raman shift. Another anomaly is found in the radial  $A_g(1)$ mode, 
which shows a downshift of  8.6 and 10.5  $cm^{-1}$ relative to  $Ba_3C_{60}$
for $K_3Ba_3C_{60}$   ($Rb_3Ba_3C_{60}$) , respectively. Such a large  
downshift upon alkali  metal doping  displays a sharp contrast  with the case 
 of $A_xC_{60}$ (A=K, Rb) where a slight upshift is observed with 
increasing  alkali metal. Low frequency $H_g$ modes also show complicated 
and characteristic  behavior. The $H_g(2)$ mode of cubic $Ba_3C_{60}$ 
is single peak at 432 $cm^{-1}$, while it shows a splitting into two peaks both
in  $K_3Ba_3C_{60}$   and $Rb_3Ba_3C_{60}$. The splitting of $H_g(3)$ 
mode is observed in all samples in Fig.2, and the splitting is smaller in the 
alkali-doped  ternaries, while the center of doublet remains unchanged.
.

Room temperature Raman spectra of  the samples  $Sm_{2.75}C_{60}$  and 
$K_{3.25}Sm_{2.75}C_{60}$ are shown in Fig.3. For 
$Sm_{2.75}C_{60}$,  an anomalously  broad distribution  of vibrational 
structures  for the low frequency  $H_g$ modes  and around the $A_g(2)$
 mode is observed, which could be related to the complexity of  
$Sm_{2.75}C_{60}$  structure.  The low  frequency  $H_g(1)$ and $H_g(2)$  
modes are asymmetric.  $H_g(2)$ mode has to be  fitted with four components.
 It suggests that the degeneracy $H_g(2)$ mode  is lifted. This splitting may be 
attributed to the symmetry lowering due to the orthorhombic  superstructure 
of this material.  A similar behavior has been  observed in single crystal 
$K_3C_{60}$  at 80 K \cite{winter} and in $Ba_4C_{60}$ and $Ba_6C_{60}$
at room temperature.\cite{chen2}   A bunch of lines appears around  700 
$cm^{-1}$ which are likely to be assigned to the $H_g(3)$ and $H_g(4)$ modes. 
It is to be  pointed out that the  pinch $A_g(2)$ mode  occurs  at 
1432.8 $cm^{-1}$, indicating  that  Sm is divalent,  and the charge transfer is
 complete according to   ~6.3 $cm^{-1}$  redshift  per electron relative to 
neutral $C_{60}$.  The  position  of the radial $A_g(1)$  is also consistent 
with that of $A_6C_{60}$ (A=K, Rb).\cite{zhou1}   This definitely
confirms divalent  Sm by Raman  scattering, being consistent with the results of 
near-edge  and extended X-ray absorption fine structure in 
$Yb_{2.75}C_{60}$.\cite{citrin}

The top of Fig.3 shows the spectrum of $K_{3.25}Sm_{2.75}C_{60}$ , 
which is amazingly   similar to $A_3Ba_3C_{60}$   although  the
Raman spectra of the intercalation host  materials  are different. 
Table I shows that  even the  positions of all  corresponding modes  
 are  almost  the same  between  $K_{3.25}Sm_{2.75}C_{60}$ and 
$A_3Ba_3C_{60}$  compounds.  This surprising  result strongly indicates  
that  Raman  spectrum is  insensitive  to the metal ion species  either
in the intercalation host  or  as an  intercalation  guest, suggesting  a
weak coupling between the  $C_{60}$ and  the metal  ions and a complete
charge transfer from metal cations to $C_{60}$ molecules. 

Raman shift of the radial $A_g(1)$ and $ A_g(2)$ pinch
modes as a function of the nominal charge transfer simply derived from the 
chemical  formula for  $Ba_3C_{60}$,   $Sm_{2.75}C_{60}$, 
$A_3Ba_3C_{60}$ (A=K, Rb), and $K_{3.25}Sm_{2.75}C_{60}$ are
plotted in Fig.4.  For comparison, the results  of 
$Ba_xC_{60}$  (x=4, 6) recently reported by us\cite{chen2}, 
  $K_xC_{60}$  (x=3, 6) reported by  Duclos  et al.\cite{duclos}, and 
$KBa_2C_{60}$  and $KBaCs_{60}$  reported by Yildirim et al.\cite{yildirim} 
are also plotted  in the figure. For the $A_g(2)$  mode, there is a boundary  
at charge transfer of  6 electrons. Below  which,  the relation between the  
Raman  shift of  $A_g(2)$  mode  and charge transfer is linear.  They follow
the characteristic shift of $\sim$6.3 $cm^{-1}$  redshift per elementary charge
for the all samples including $Ba_3C_{60}$ and $Sm_{2.75}C_{60}$. However,
when the charge transfer exceeds six,  the pinch $A_g(2)$  mode  does not follow
the simple  characteristic  relation, showing a complicated behavior . In other
words, although there exist a systematic relation between Raman shift and charge
transfer for the "$t_{1u}$  band" fullerides, the  "$t_{1g}$ band"  fullerides
are quite different. The charge transferred from metal to $C_{60}$ 
molecules  were  -9, -15, and -7 for the $Ba_4C_{60}$, $Ba_6C_{60}$,  and 
$K_3Ba_3C_{60}$ ( $K_{3.25}Sm_{2.75}C_{60}$ ), respectively, if charge 
transfer was derived according to ~6.3$cm^{-1}$  redshift per elementary charge. 
This is much larger ( less ) than the nominal charge transfer  for $Ba_xC_{60}$ 
(x=4 and 6) ($K_3Ba_3C_{60}$ and $K_{3.25}Sm_{2.75}C_{60}$). 
If the charge transfer from Ba to $C_{60}$  molecules  is complete, the  
redshift  per elementary charge is  about  9 $cm^{-1}$, being  much larger than  
$\sim$6.3 $cm^{-1}$.  In addition,  an anomalous  Raman shift  for the $A_g$  
derived modes  takes place  in the ternary   $A_3Ba_3C_{60}$ (A=K, Rb) and 
$K_{3.25}Sm_{2.75}C_{60}$, the down-shift  is only  about  6 $cm^{-1}
$ for the nominal charge  transfer of three electrons  relative to $Ba_3C_{60}$  
and $Sm_{2.75}C_{60}$.

Let us switch to the arguments on the radial  $A_g(1)$ mode.  A continuous
 up-shift of the $ A_g(1)$ mode with increasing  charge transfer until six
electrons observed as shown in Fig.4. This mode-stiffening  has been explained
by electrostatic interactions which produces  sufficient stiffening to encounter 
the softening of the mode  expected  on the basis of charge transfer effects.
\cite{jishi,chen} It is to be noted that  the Raman shift of  the radial $A_g(1)$ 
for  $Sm_{2.75}C_{60}$  with $t_{1u}$ states  falls on the linear  line of
 $K_xC_{60}$ with $t_{1u}$ states. In contrast , in the $t_{1g}$  fullerides, 
the  frequency  of the  $A_g(1)$ mode  almost  remains  unchanged for 
$Ba_xC_{60}$,  while it decreases  with increasing charge transfer  in the 
ternaries  $A_3Ba_3C_{60}$  and  $K_{3.25}Sm_{2.75}C_{60}$. It is worth 
noting  that both $A_g(1)$  and $A_g(2)$ modes show complicated  behavior 
in the $t_{1g}$ bands.

In the case of Ba derived fullerides, a strong hybridization  between the Ba
atoms and the $\pi$-type functions of the $C_{60}$  network
\cite{saito,erwin,niedrig} may be responsible for the behavior of the $A_g$
derived modes as discussed in Ref.15. The charge transfer is complete  in the
intercalation hosts  of $Ba_3C_{60}$ and $Sm_{2.75}C_{60}$, following the
characteristic relation between Raman shift and charge transfer, while the
introduction of  alkali metals  leads to an anomalous  Raman shift for the
two $A_g$ derived modes  in the  ternary  $A_3Ba_3C_{60}$ and
$K_{3.25}Sm_{2.75}C_{60}$,. However, the Raman shift of $A_g(2)$  pinch 
mode  follows the  characteristic  shift of $\sim$6.3 $cm^{-1}$ per elementary 
charge transfer  for  the co-intercalated  $C_{60}$ of  $ABa_2C_{60}$ (A=K, Rb,
and Cs) and $KBaCsC_{60}$ with $t_{1u}$ states.\cite{yildirim}   It suggests
that the anomalous  Raman  shift of the $A_g$ mode could be related to the
"$t_{1g}$" band  electrons. In fact, some other differences between  the
"$t_{1u}$" and "$t_{1g}$" superconductors have been found. In the $t_{1u}$
band, superconductivity  occurs only at $(C_{60})^{-3}$ state 
which corresponds to the half-filling of the  $t_{1u}$  band.
\cite{yildirim1} However, the reported "$t_{1g}$" superconductors are
$B_4C_{60}$ (B=Ba, Sr), $A_3Ba_3C_{60}$, and $Ca_5C_{60}$. which have the 
molecular valences of  -8, -9,  and -10, respectively. Such tolerance for the
molecular valence in $t_{1g}$ superconductor makes a striking contrast with the
strict constraint for the valence state in the case of "$t_{1u}$"
superconductors.  Another difference to be pointed out is that the "$t_{1u}$"
superconductivity appears only in fcc or related structures. In the case of
$t_{1g}$ band, superconductivity is observed in various structures, even
orthorhombic structure.  To completely understand  the physical properties
and superconducting  mechanism of fullerides, it is necessary to further
investigate  $t_{1g}$ fullerides, and to explain the difference between
$t_{1u}$   and $t_{1g}$ fullerides.

In summary,  the Raman  scattering  study  has  been carried out in the  two 
fulleride families   of  $A_xBa_3C_{60}$ (x=0, 3; A=K, Rb) and 
$K_xSm_{2.75}C_{60}$  (x=0, 3.25). The results definitely show some 
differences  between $t_{1u}$ and $t_{1g}$  fulleride  for Raman shift of the 
two $A_g$ modes. For $t_{1u}$ fullerides, the down-shift  of the  $A_g(2)$ mode 
and the up-shift of $A_g(1)$ mode  can be explained  by charge  transfer and the 
electrostatic  interactions  upon doping, respectively.  In contrast,  the $t_{1g}$
fullerides  show  a complicated behavior  for Raman shift of the two $A_g$ modes.  
The  Raman  shifts of the two $A_g$ modes , observed in the tenaries   
$A_3Ba_3C_{60}$ and $K_{3.25}Sm_{2.75}C_{60}$, significantly  differ 
from that of the binary fullerides $Ba_4C_{60}$ and $Ba_6C_{60}$. It is also 
found that  Raman spectra of  $A_3Ba_3C_{60}$  and 
$K_{3.25}Sm_{2.75}C_{60}$ are  identical to each other, indicating  a weak  
coupling between the $C_{60}$ and  metal cations,  and the same charge transfer 
 to $C_{60}$ molecules. However,  the anomalous  Raman shift could  not be 
explained  only by the charge transfer.

\vspace{5mm}

X. H. Chen would like to thank the Inoue Foundation for Science  for financial
support. X. H. Chen acknowledges   Dr. Kitagawa for his help in experiment.
This work is partly supported by  Grant from the Japan Society  for 
Promotion of Science  (RFTF 96P00104, MPCR-363/96-03262) and from 
the Ministry of Education, Science, Sports, and Culture.

\begin{table}
\caption{ Positions and linewidths  (in parentheses)  for the Raman modes  in
$Ba_3C60_{60}$,  $A_3Ba_3C_{60}$  (A=K and Rb), $Sm_{2.75}C_{60}$, and 
$K_{3.25}Sm_{2.75}C_{60}$ }
\renewcommand{\arraystretch}{0.93}
\begin{tabular}{c c c c c c} 
         &  $Ba_3C_{60}$   &   $K_3Ba_3C_{60}$   &  $Rb_3Ba_3C_{60}$   &  
     $Sm_{2.75}C_{60}$  &  $K_{3.25}Sm_{2.75}C_{60}$  \\
$I_h$  mode   &  $\omega$   ( $\gamma$  )  &  $\omega$   ( $\gamma$  )  &  
$\omega$  ( $\gamma$ )   &   $\omega$   ( $\gamma$ )  &  $\omega$  ( $\gamma$ )  \\
        &     ( $cm^{-1}$  )   &  ( $cm^{-1}$  )   &   ( $cm^{-1}$  )   &
        ( $cm^{-1}$  )   &  ( $cm^{-1}$  )   \\ \hline
$A_g(1)$    &   505.9 ( 4.2 ) &  497.3( 2.4 )  &  495.4 ( 3.7 ) &  498.3 ( 6.9 ) 
   &  496.8 ( 5.0 )  \\
$A_g(2)$    &   1430.8 ( 13.0 ) &  1425.5 ( 6.3 )  &  1424.1 ( 11.1 )  
&   1432.8 ( 22.6 ) &  1424.7 ( 14.3 ) \\
$H_g(1)$   &   273 ( 5.3 )   &   266.3 ( 11.9 )  &  266.6 ( 4.5 ) 
&   264.7 ( 23.3 )  &  265.6 ( 10.4 )     \\
                   &   278.7 ( 4.6 )     &   276.5 ( 3.0 )   &  272.8 ( 5.7 )
&  283.0 ( 3.1 )  & 276.0 ( 6.6 ) \\
$H_g(2)$   &   432.3 ( 5.3 )    &   414.9 ( 2.3 ) &  417.8 ( 3.7 )    
  &  344.5 ( 38.3 )  &  415.0 ( 3.9 ) \\
      &        &   422.9 ( 2.8 )  &  424.8 ( 2.9 ) &  
    400 ( 48.8 ) & 422.7 ( 18.0 ) \\    
    &          &         &      &   415.0 ( 18.0 )  &      \\
   &           &         &      &   427.0 ( 10.2 )    &     \\
$H_g(3)$  &   648.2 ( 8.5 )  &   653.5 ( 5.6 ) &   655.3 ( 5.8 ) & 
647.9  (  25.0 )  &  653.3 ( 7.2 )  \\
                  &     681.6 ( 7.8 )  &   673.3 ( 4.8 )  &  674.3 ( 4.2 )
 &   683.4 ( 31.0 ) &  672.1 ( 12.9 )    \\
$H_g(4)$  &   760.7 ( 8.4 )  &   757.7 ( 4.2 ) & 758.7 ( 5.6 )    
   &   756.7 ( 23.0)  & 757.5 ( 7.8 )      \\
$H_g(5)$  &  1091.7 ( 18.5)   &   1086.0 ( 15.3 )  &  1086.7 ( 16.4 ) 
  &  1085.5 ( 22.3 ) & 1088.0 ( 25.2 )       \\
                  &   1117.3 ( 12.8 )  &     &        & 1112.4 ( 12.8 )   &    \\
$H_g(6)$  &   1227.6 ( 16.1)  &  1229.0 ( 9.8 ) &  1227.9 ( 13.1 )    
&  1226.2 ( 20.9 )  &  1227.6 ( 16.7 )        \\
$H_g(7)$  &   1322.1 (34.6)   &     1314.0 ( 20.6 )   & 1315.5 ( 18.0 )    
 &         & 1317.9  ( 27.1 )     \\
                  & 1388.1 ( 15.3 ) &     1377.4 ( 8.9 )   & 1373 ( 23.5 )    
  &     1387 .8 ( 46 ) & 1378.2 ( 11.5 )      \\
$H_g(8)$  &    1474.4 ( 26.1 ) &     &     &    &        \\
\end{tabular}
\end{table}

\noindent
{\bf FIGURE CAPTIONS} \\

\noindent
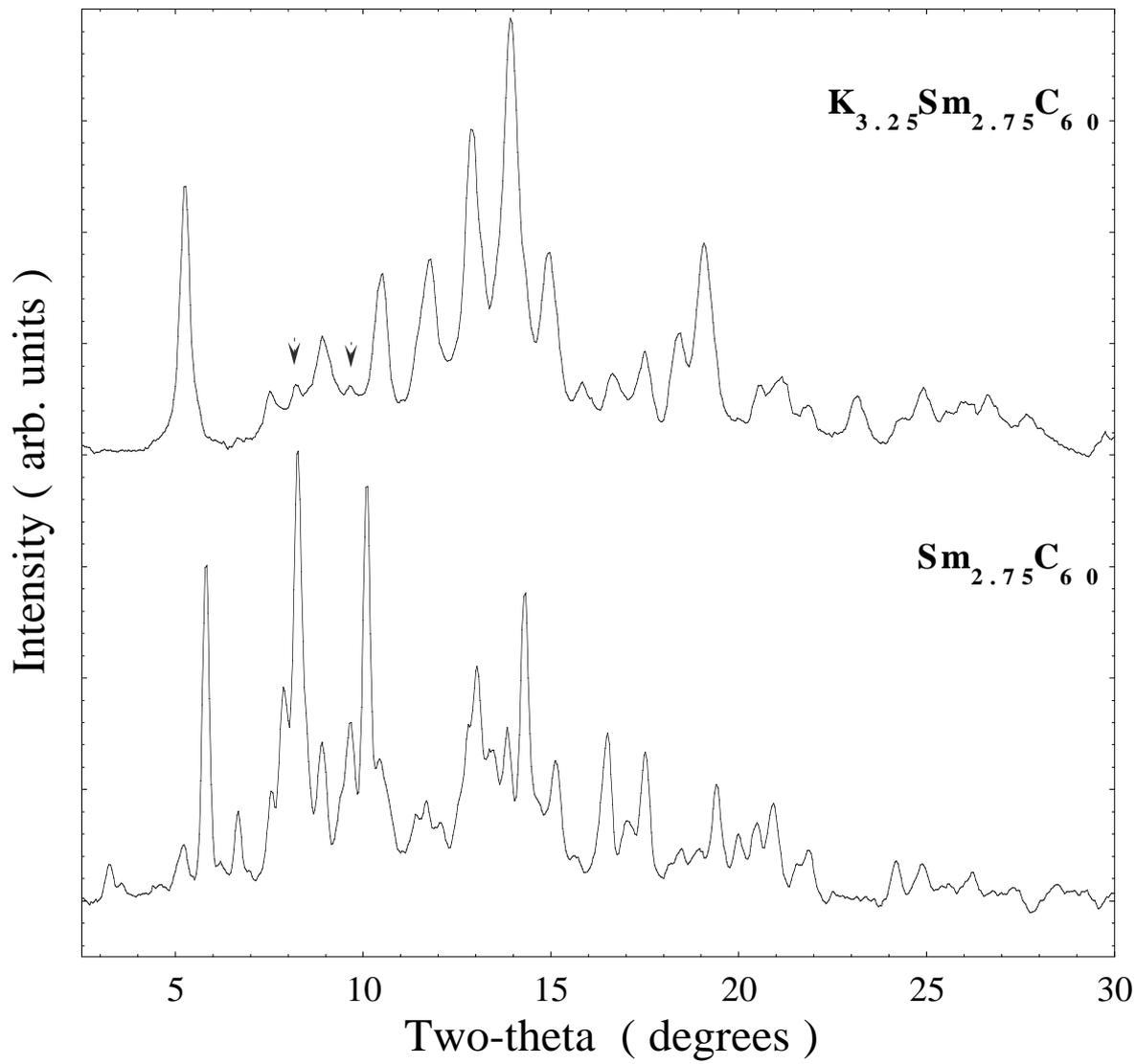
Figure 1:

Powder  X-ray diffraction  patterns  of $Sm_{2.75}C_{60}$ and 
$K_{3.25}Sm_{2.75}C_{60}$. The two peaks marked by arrows can not 
indexed with body-centered-cubic cell with  lattice parameter a=11.093\ \r{A}.\\

\noindent
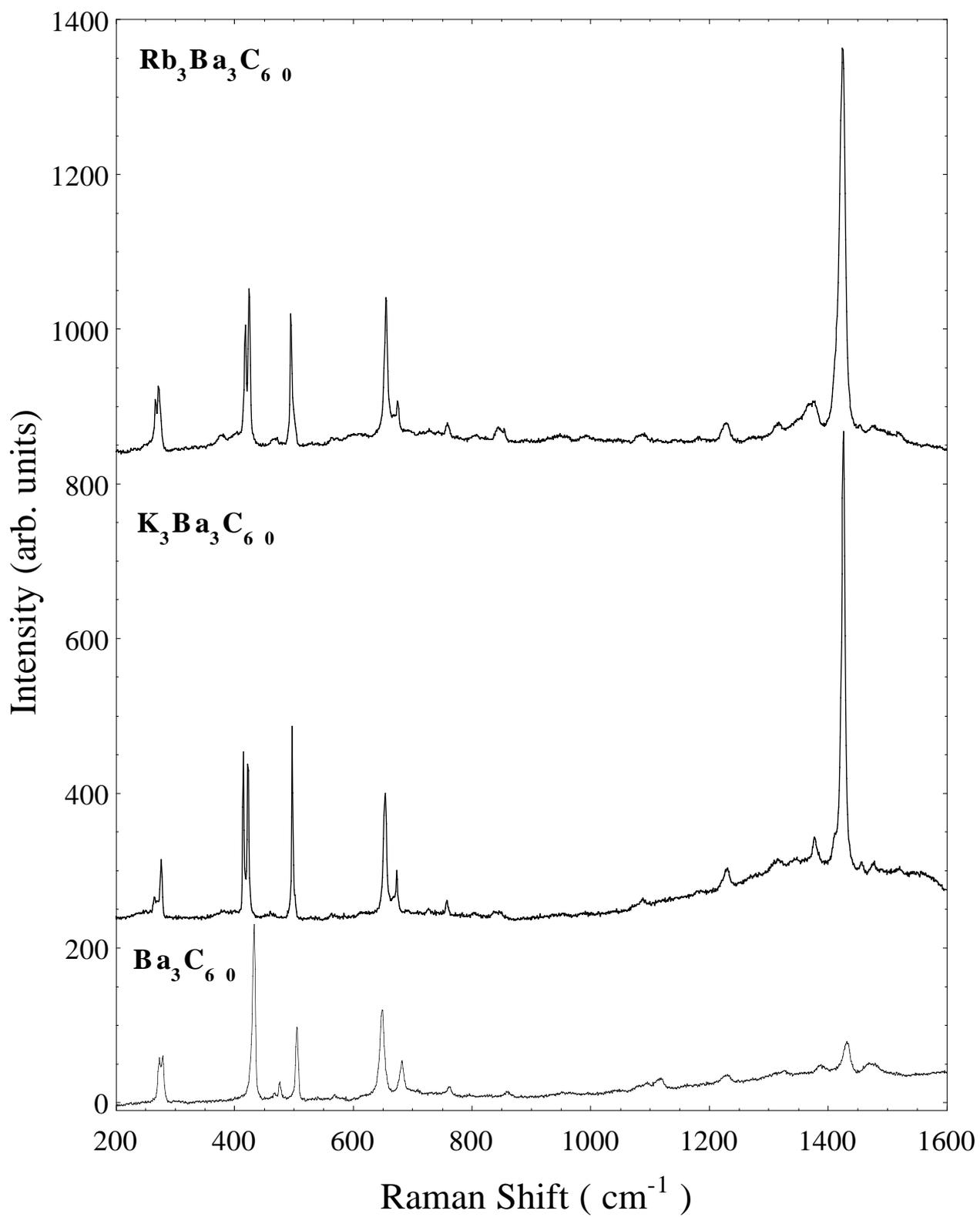
Figure 2:

Room temperature  Raman spectra of  $Ba_3C_{60}$,  $K_3Ba_3C_{60}$,  
and  $Rb_3Ba_3C_{60}$. \\

\noindent
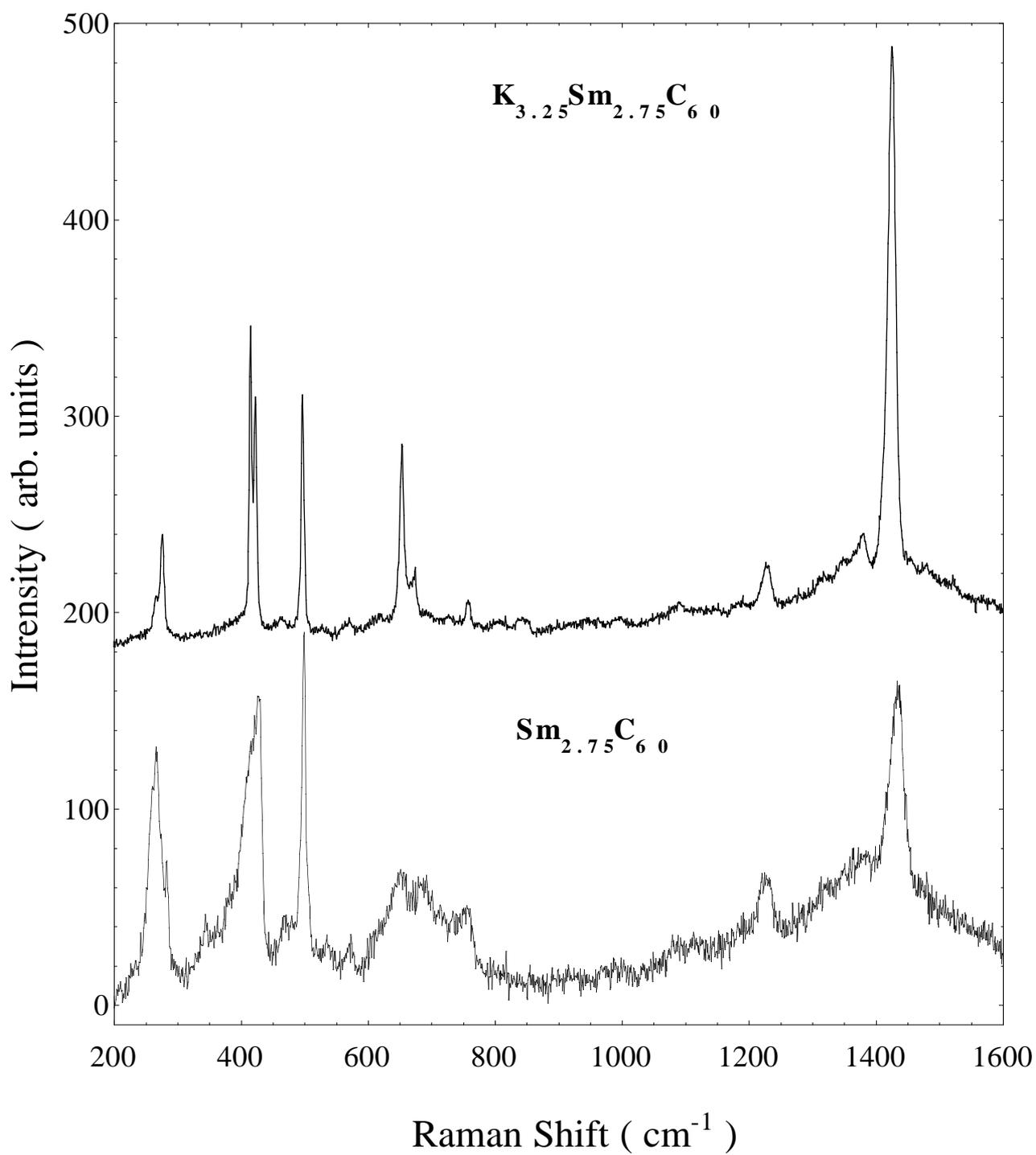
Figure 3:

Room temperature Raman spectra of  $Sm_{2.75}C_{60}$ and
$K_{3.25}Sm_{2.75}C_{60}$.\\

\noindent
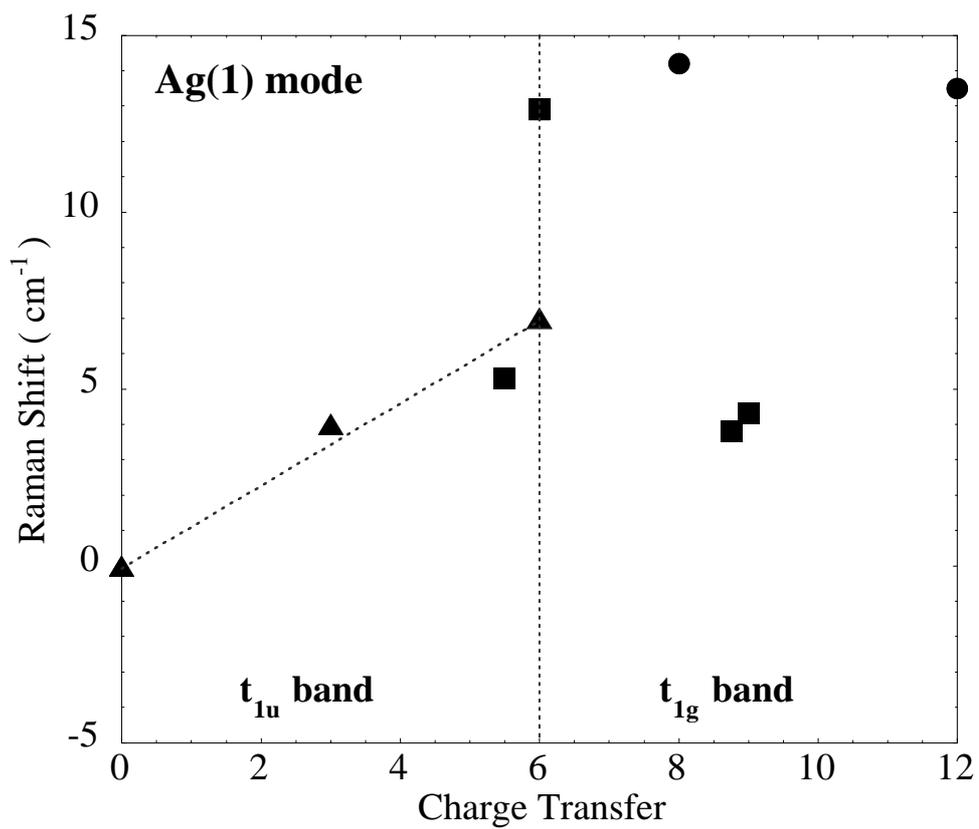
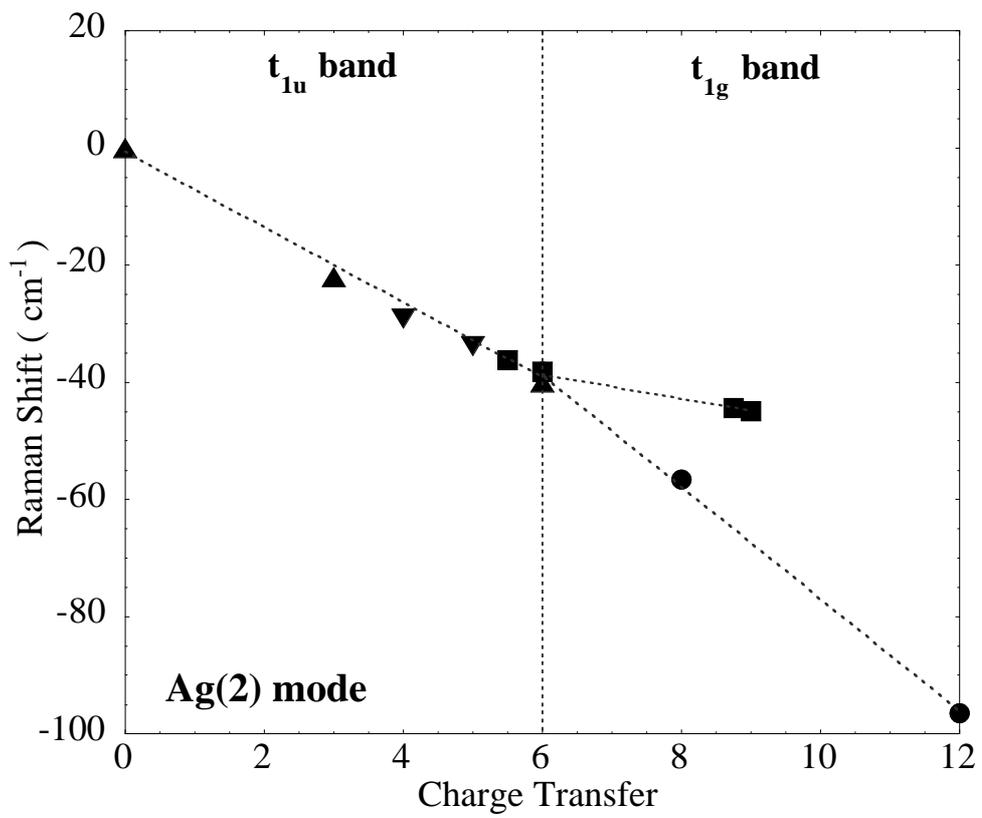
Figure 4:

Charge  transfer-Raman shift relation for the  radial $A_g(1)$ and $A_g(2)$ 
pinch modes. Squares  represent  the current  experimental  results,  circles  
refer to the experimental   results of $Ba_xC_{60}$  reported by us (Ref.15),
up-triangles  are   from the results  of $K_xC_{60}$  reported  by  
Duclos  et al. (Ref.5), and down-triangles  refer to the results of 
$KBa_2C_{60}$ and $KBaCsC_{60}$ reported by Yildirim et al. (Ref.16).


\begin{references}

\bibitem{mitch}
M. G. Mitch, S. J. Chase and Lannin, Phys. Rev. Lett. {\bf 68}, 883(1992).


\bibitem{zhou}
P. Zhou, K. A. Wang, P.C. Eklund, G. Dresselhaus, M.S. Dresselhaus, Phys. 
Rev. B {\bf 48}, 8412(1993).

\bibitem{winter}
J. Winter and H. Kuzmany  Phys. Rev. B {\bf 53}, 655(1996).


\bibitem{haddon}
R.C. Haddon, A.F. Hebbard, M.J. Rosseinsky, D.W. Murphy, A.J. Duclos, 
K.B. Lyons, B. Miller, J.M. Rosamilia, R.H. Fleming, A.R. Kortan, 
S.H. Glarum, A.V. Makhga, A.J. Muller, R.H. Eick, S.M. Zahurak, 
R.Tycko, G. Dabbah, and F. A. Thiel, Nature {\bf 350}, 320(1991).


\bibitem{duclos}
S. J. Duclos, R. C. Haddon, S. H. Glarum, A. F. Hebbard and K. B. Lyons,
Science {\bf 254}, 1625(1991).

\bibitem{pichler}
T. Pichler, M.Matus, J.K\"{u}rti, and H. Kuzmany, Phys. Rev. B {\bf45},
13841(1992).

\bibitem{wang}
K.A. Wang, Y. Wang, P. Zhopu, J.M. Holden, S.L. Ren, G.T. Hager, 
H.F. Ni, P.C. Eklund, G. Dresselhaus, and M.S. Dresselhaus.
Phys. Rev. B {\bf45},  1955(1992).

\bibitem{kuzmany}
H. Kuzmany , M. Matus, B. Burger, and J. Winter,  Adv. Mater. 6, 731(1994).

\bibitem{winter1}
J. Winter and H. Kuzmany, Solid State Commun. {\bf 84} 935(1992).

\bibitem{iwasa}
Y. Iwasa, H. Hayashi, T. Furudate, and T. Mitani, Phys. Rev. B
{\bf 54}, 14960(1996).

\bibitem{iwasa1}
Y. Iwasa, M. Kawaguchi, H. Iwasaki, T. Mitani, N. Wada, and T. Hasegawa
 Phys. Rev. B {\bf 57}, 13395(1998).

\bibitem{chen}
X.H. Chen, X. J. Zhou, and S. Roth, Phys. Rev. B {\bf 54}, 3971(1996).

\bibitem{chen1}
X.H. Chen and G. Roth, Phys. Rev. B {\bf 52}, 15534(1995).

\bibitem{zhou1}
P. Zhou, K. A. Wang, Y. Wang, P. C. Eklund, M. S. Dresselhaus, 
G. Dresselhaus, R. A. Jishi Phys. Rev. B  {\bf 46}, 2595(1992).

\bibitem{chen2}
X.H. Chen, S. Taga, and Y. Iwasa, submitted to Phys. Rev. B

\bibitem{yildirim}
T. Yildirim, L. Barbedette, J.E. Fisher, G.M. Bendele, P.W. Stephens
 C.L. Lin, J. Robbert, P. Petit, and T.T.M. Palstra,
Phys. Rev. B {\bf54}, 11981(1996).

\bibitem{citrin}
P.H.  Citrin, E. \"{O}zdas, S. Schuppler , A.R. Kortan, and K.B. Lyons
Phys. Rev. B {\bf56}, 5213(1997).

\bibitem{jishi}
R.A. Jishi and M.S. Dresselhauss, Phys. Rev. B {\bf45}, 6914(1992).

\bibitem{saito}
S. Saito and A. Oshiyama, Phys. Rev. Lett. {\bf71}, 121(1993).

\bibitem{erwin}
S.C. Erwin and M.P. Pederson, Phys. Rev. B {\bf47}, 14657(1993).

\bibitem{niedrig}
Th.S. Niederig, M.C. B\"{o}hm, H. Werner, J. Schlute, and R. Schl\"{o}gl,
Phys. rev. B {\bf55}, 13542(1997).


\bibitem{yildirim1}
T. Yildirim, L. Barbedette, J.E. Fisher, C.L. Lin, J. Robbert, P. Petit, 
and T.T.M. Palstra, Phys. Rev. Lett. {\bf77}, 167(1996).

\end{references}
\end{document}